\documentclass[a4paper,twocolumn]{esapub} 
\usepackage[dvips]{graphicx}
\usepackage{natbib}
\bibpunct{(}{)}{;}{a}{}{;}
\usepackage{times}
\title{ Oscillations and Waves in coronal loops$^1$}
\author{Tongjiang Wang}

\affil{Max-Planck-Institut f\"ur Sonnensystemforschung, 37191 Katlenburg-Lindau,
Germany\\
email:  wangtj@mps.mpg.de }

\newcommand{\kms}{km~s$^{-1}$}
\begin{document}

\maketitle

\footnote{In Proc. of {\em Chromospheric and Coronal Magnetic Fields, Aug.30-Sep.2, 2005,
in MPS, Katlenburg-Lindau, Germany }, ed. D. Danesy (ESA SP-596)}


\begin{abstract}
In the past few years observations by high-resolution space imaging telescopes
and spectrometers have confirmed that a great variety of MHD waves are supported
in the solar corona of a low-beta plasma and fine structure. MHD waves are an
important diagnostic tool for the determination of the physical parameters of
coronal loops, dubbed {\em coronal seismology}. In this paper, I will review recent
results of both propagating and standing waves observed with SOHO and TRACE,
and discuss the wave damping and excitation mechanisms as well as some
applications of coronal seismology based on recent numerical simulations and
theories in relation to the observations.
\end{abstract}

\section{Introduction}
Recent observations of oscillations and waves in coronal loops have inspired a
great deal of interest in {\em coronal seismology}
\citep[see the reviews by][]{rob00, nak04a}, which was proposed
by B. Roberts more than 20 years ago \citep{rob84}. The coronal seismology,
as a diagnostic tool, opens the possibility for uncovering the physical properties
of the coronal structure by studying how they influence the waves in excitation,
propagation and dissipation. These properties include the temperature and density
structure, the magnetic field structure, and the magnitudes of important transport
coefficients. Constrains on dissipation coefficients and the strength of the
coronal magnetic field may resolve some of the existing difficulties with
wave heating and reconnection theories.

\begin{figure}
\centering
\includegraphics[width=0.8\linewidth]{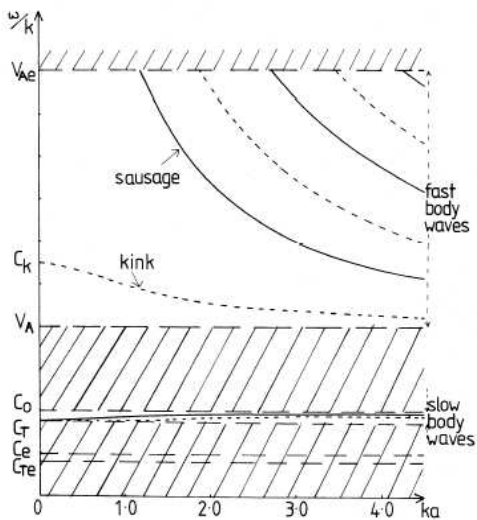}
\caption{\label{dispers}
 The phase-speed as a function of longitudinal wavenumber for the fast and slow
magnetoacoustic waves in a flux tube under coronal conditions,
$v_{Ae}>v_{A}>c_0>c_T>c_e$. {\em Solid curves}, sausage modes; {\em dashed curves},
kink modes.  \citep[excerpted from][]{edw83} }
\end{figure}

Edwin \& Roberts (1983) predicted that various kinds of MHD
waves and oscillations can be supported by coronal loops. They derived
dispersion relation for the oscillations of a straight magnetic cylinder
under coronal conditions. Figure~\ref{dispers} shows three kinds of free 
mode oscillations: slow modes, fast modes and torsional Alfv\'{e}n modes. Slow modes 
have phase speeds close to the sound speed. Fast modes include two solutions,
termed sausage modes and  kink modes, with the oscillations
symmetrical and asymmetrical about the axis of the loop, respectively. 
Note that the condition of the Alfv\'{e}n speed inside the loop lower than that
in its surroundings implies that a dense coronal loop is necessary to act as wave 
ducts trapping fast mode waves. Torsional Alfv\'{e}n modes correspond to pure Alfven 
waves in the cylindrical case. Patterns of the different modes were sketched out
by \citet{wan04} (see Fig.1 in his paper), which can be used for mode identifications
in observation. Slow sausage modes are dominated by longitudinal oscillations with 
magnetic pressure and thermal pressure perturbations coupling in anti-phase.
While fast sausage modes are dominated by transverse oscillations in the radial 
direction with the two restoring forces in phase. Fast kink mode oscillations appear as
the lateral displacements, with a small density perturbation approximating to
be zero in the first order. Torsional Alfv\'{e}n modes show periodic spinning motions
about the axis with no density perturbation. 

In observations these modes may be distinguished. For example, since the slow and fast 
sausage modes cause the density and magnetic field variations, they will modulate plasma
emission (e.g., free-free emission, atomic line emission, or bremsstrahlung) 
and radio emissions (driven by a gyro-synchrotron mechanism or loss-cone instability).
\citet{zaq03} suggests that the global torsional Alfv\'{e}n oscillation of coronal 
loops may be observed by the periodic variation of a spectral line width.
These modes are also distinctly different in their periods.  
\citet{asc03a} estimated for typical coronal loops (with lengths of 50$-$500 Mm) 
that the slow modes have periods in a
range from 7 to 70 minutes, the fast kink modes have periods from several minutes to 
14 minutes, and the fast sausage modes have periods on the order of seconds.

\begin{figure*}
\centering
\includegraphics[width=0.8\linewidth]{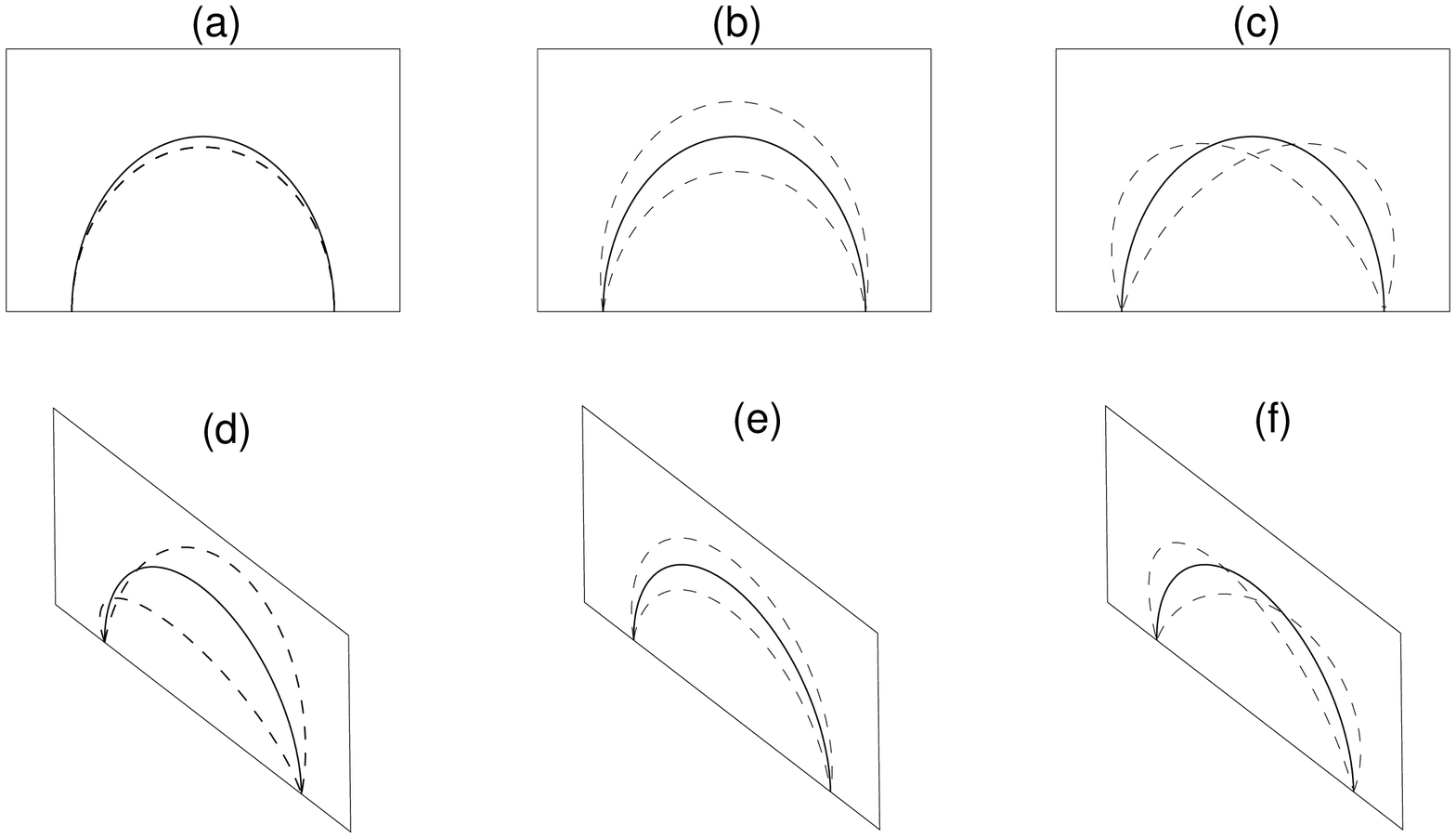}
\caption{\label{kink}
  Demonstration of three kinds of global kink mode loop oscillations. (a) horizontal
loop oscillation, (b) vertical loop oscillation, and (c) distortion loop oscillation.
(d)-(f) Same as (a)-(c) but viewed from the top-left.}
\end{figure*}

For a straight flux tube the kink mode oscillations in any direction are identical, 
whereas for a line-tied coronal loop we may need to distinguish three kinds of global
kink oscillations (see Fig.~\ref{kink}). The first, {\em horizontal oscillation}, shows 
swaying motions perpendicular to the loop plane.  The second, {\em vertical oscillation},
is polarized in the loop plane, showing expanding and shrinking motions. 
For the third, {\em distortion oscillation}, the loop gets distorted in the loop plane 
while nearly remaining its length. The horizontal and vertical loop oscillations have 
been identified in observation. The distortion modes can be generated in 2D simulations.

In the previous studies, the existence of coronal oscillations had been inferred 
from time profiles of quasi-periodic patterns observed in almost all wavelengths, 
but the majority in radio wavelengths \citep[see reviews by][]{asc87, asc03a}.
Recent imaging observations, especially from SOHO and TRACE in EUV wavelength with 
high spatial and temporal resolutions allow us to identify the oscillation
modes unambiguously. For example, both the horizontal and vertical kink-mode loop
oscillations have been observed by TRACE in EUV \citep[e.g.][]{asc99, nak99, wan04b}.
The global standing slow-mode oscillations of hot coronal loops have been observed
by SUMER on SOHO \citep[e.g.][]{wan02, wan03a}. The quasi-periodic pulsations were
found in radio wavelengths and ascribed to global sausage modes \citep{nak03}. Not only
the standing waves but also propagating waves have been found in coronal loops.
The propagating slow magneto-acoustic waves in polar plumes \citep{def98} and 
long fan-like coronal loops \citep[e.g.][]{ber99, dem00} were detected by SOHO/EIT 
and TRACE in EUV. The fast sausage mode waves were found by SECIS in the coronal
green line of Fe XIV~\citep{wil01, wil02}, and the propagating fast kink waves in 
the post-flare supra-arcade by TRACE \citep{ver05}. For a detail, refer to the recent 
observational reviews given by \citet{wan04, asc04}

In this paper, I will review the recent results from SOHO and TRACE and their 
interpretations.

\section{Global fast kink-mode oscillations}

\subsection{Observations of Horizontal and Vertical Loop Oscillations}

Using TRACE observations, \citet{asc99} first discovered the horizontal 
oscillations of active region loops. After a strong M-class flare at 12:55 UT 
on 1998 July 14, a number of at least five loops in the flaring active region 
exhibited periodic horizontal displacements. Almost all loops initially moving 
away from the central flare site suggest that the oscillations could be excited 
by a blast shock wave, as also evidenced by a propagating disturbance with a 
speed of about 700 \kms.  \citet{sch02} suggested that the oscillations 
also could be excited by the ejected plasma in CMEs and filament eruptions 
directly impacting on the loop. Their ideas are supported by a statistic study
by \citet{hud04} that 12 of 28 oscillation cases were
associated with metric type II bursts and 24 of 28 cases associated with CMEs.

\begin{figure*}
\centering
\includegraphics[width=0.8\linewidth]{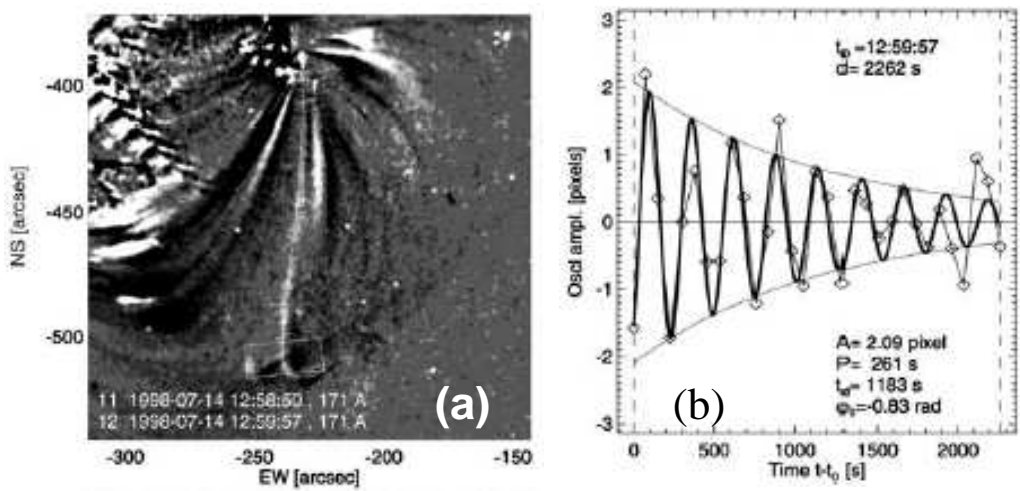}
\caption{\label{horizon}
Horizontal loop Oscillation event on 1998-Jul-14, 12:45 UT. (a) Difference image with
rectangular box indicates where the oscillations are analyzed. (b) Detrended
oscillation, fitted with an exponentially decaying oscillatory function.
\citep[from][]{asc02} }
 \end{figure*}

\citet{asc02} measured geometric and physical parameters of transverse oscillations
in 26 coronal loops, out of the 17 events described in \citet{sch02}.
Figure~\ref{horizon} demonstrates an example for measurements of the spatial
displacements of an oscillating loop. The time profile shows that the oscillation
is strongly damped. They obtained the displacement amplitude, period, and
decay time by fitting with an exponentially damped sine function. They find that
oscillation periods have a mean value of $P=5.4\pm2.3$ min, covering a range of
$P=2.3-10.8$ min. The decay times have a mean of $t_d=9.7\pm6.4$ min, in a range of
$t_d=3.2-21$ min, and there are 9 cases for which the decay time could not be
evaluated. The transverse amplitudes are on average $2200\pm2800$ km, with the
maximum of 8800 km. The maximum transverse velocities are $42\pm53$ \kms, up to
230 \kms.  From the measured loop length with a mean value of $L=220$ Mm and 
the oscillation period, the phase speed for the fundamental mode, $c_k=2L/P\simeq1400$
\kms, slightly larger than the typical Alfven speed ($v_A\simeq1000$ \kms) in the corona,
consistent with the interpretation in terms of the global fast kink mode oscillations.

\begin{figure}
\centering
\includegraphics[width=0.98\linewidth]{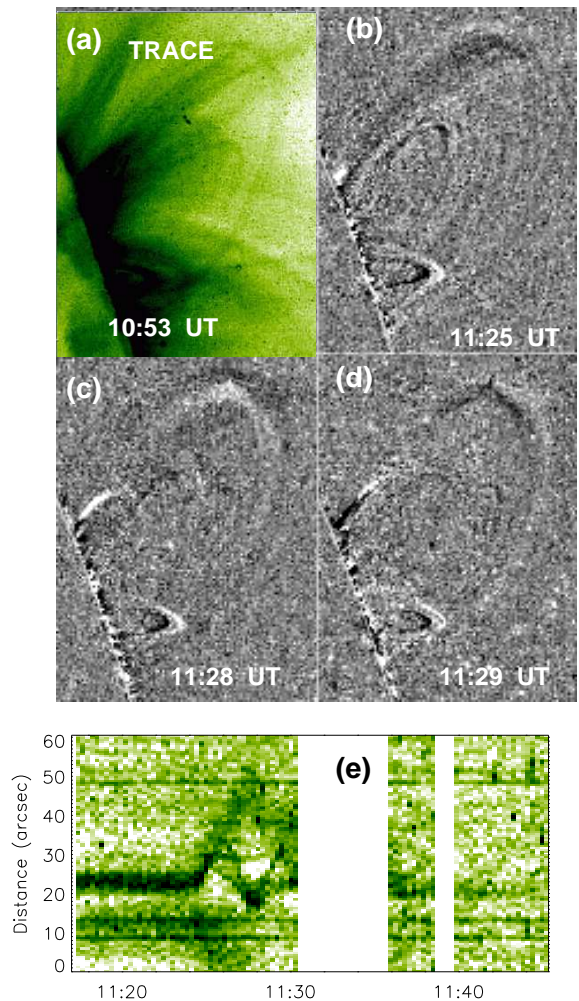}
 \caption{\label{vertic}
  (a) A TRACE 195\AA~image containing the analyzed coronal loop. The box marks
the location of a cut, whose time series is shown in (e). (b)-(d) Running difference
images with an interval of $\sim$2 min. Black indicates where the loop was
in the earlier image and white where it has moved to. (e) Time slices of TRACE intensity
along a cut at the loop apex. \citep[from][]{wan04b} }
\end{figure}

\citet{wan04b} recently found the first evidence for vertical loop oscillations.
In order to identify the oscillation mode they simulated the horizontal and vertical
oscillations in a synthetic loop with a geometry derived from the observation,
and found that they display distinctly different signatures.
The horizontal oscillation exhibits a cross-over between the black and white
shading near the loop-top, whereas the vertical oscillation appears as a uniform
displacement with a maximum contrast at that position. In observations 
(see Figs.~\ref{vertic}b-d), the signature of white loop either inside or outside 
the black one is consistent with that of  the vertical loop oscillation.
Time series of a slice at the loop top reveals intensity variations of the loop
(Fig.~\ref{vertic}e), which can be explained by the loop length modulations 
during the oscillation, and also suggest that the vertical loop oscillation could
be more compressible than the horizontal loop oscillation.

\subsection{Excitation of Global Kink Mode Oscillations}

However, the detected transverse loop oscillations in flares are relatively rare.
A statistic study by \citet{sch02} showed that in only 17 events (about 6\%)
out of 255 flares inspected,  the oscillations were excited. This implies that some
special excitation conditions are required. A criterion for the excitement
of oscillations suggested by \citet{asc02} is the need of enough fast exciter
velocity, as inferred from the fact that the majority of oscillation events
(about 70\%) are M- or X-class flares. They also suggested that loops with weaker
magnetic fields have a higher likelihood to oscillate than loops with strong
magnetic fields, because loops with lower magnetic fields are excited faster than
they can restore the equilibrium, and thus overshoot and swing back. The other
criterion suggested by \citet{sch00, sch02} is the location of oscillatory loops
near magnetic nullpoints or separators, because a little disturbance will be
highly amplified at these places.

In theory, to understand the excitation and damping mechanisms of the observed 
fast kink-mode oscillations, 1D \citep{ter05}, 2D \citep{sel05a, mur05, del05},
and 3D simulations \citep{miy04} have been tested.  However, since the equilibrium 
of a curved loop with enhanced density under a realistic condition (e.g.  with the
gravitational and temperature stratifications) is difficult to set up, these
simulations are still far from reproducing the observations.

\subsection{ Damping Mechanism of Kink Mode Oscillation}

Various kinds of damping mechanisms have been proposed to explain the rapid
decay of transverse loop oscillations observed by TRACE, such as footpoint 
motions near separatrices \citep{sch00}, footpoint wave leakage into the 
chromosphere \citep{dep01, ofm02a}, Lateral wave leakage due to curvature 
of loops \citep{rob00}, phase mixing \citep{hey83, nak99, ofm02b}, resonant 
absorption \citep{rud02, goo02, asc03b, van04a}, non-ideal MHD effects 
(e.g., viscous and 
Ohmic damping, optically-thin radiation, thermal conduction) \citep{rob00}.  
Observational tests based on physical parameters measured in 11 oscillation
events indicate that the damping of fast kink-mode waves are most likely due to 
phase mixing or resonant absorption involved in the small-scale inhomogeneity.

If an observed loop consists of multiple unresolved loop threads,
the oscillations are damped by phase mixing because different threads have
different oscillation frequencies which lead to enhanced viscous and ohmic
dissipations by the friction between adjacent threads.
Assuming the inhomogeneity scale is proportional to the loop length or the loop
width, \citet{ofm02b} found that the scaling power of the damping time with the
observed parameters of the loops agrees well with the power expected by phase 
mixing, and they obtained the anomalously high viscosity, about 8$-$9 orders of magnitude
larger than the classical value.

Another mechanism, resonant absorption interprets the rapid damping of loop
oscillations by a mode conversion process, i.e., global kink
mode oscillations transfer energy into torsional Alfven modes of the
inhomogeneous layers at the loop boundary. Since this mechanism is not
directly involved in the wave dissipation, it does not need to invoke
the enhanced viscosity. Observational test by \citet{asc03b}
supports the resonant absorption as a possible damping mechanism, and
a new diagnostic of the density contrast of coronal loops can be provided
based on this theory.

Some recent work in theory and numerical simulation also explore the effects
of loop curvature \citep{van04b, bra05, sel05a, mur05} and gravitational stratification 
on the damping \citep{miy04, and05a, del05}.

\section{Global slow mode oscillations}

\subsection{Observations of Hot Loop Oscillations}

\begin{figure}
\centering
\includegraphics[width=0.9\linewidth]{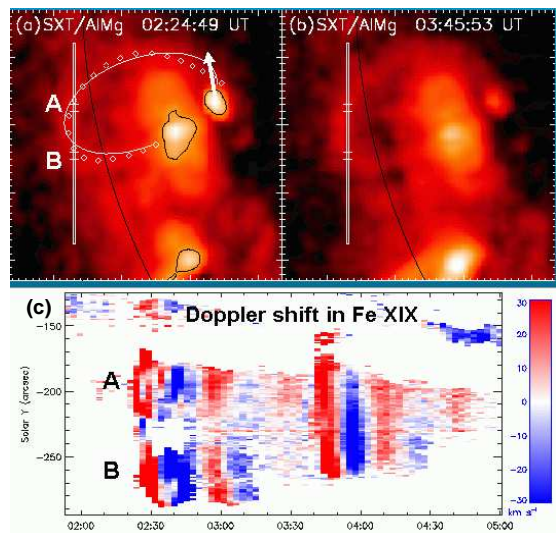}
\caption{\label{sxrdp}
  (a) The oscillating soft X-ray loop (outlined with $diamonds$)
fitted with an elliptical model (white curve) for the 29 September 2000
events. The SUMER spectrometer slit position is indicated.
(b) The SXT image at 03:45:53~UT. (c) Doppler shift time series in the Fe XIX line.
The redshift is represented by red color, and the blueshift by blue color.
 \citep[from][]{wan03b} }
 \end{figure}

\citet{wan02} found the first evidence for global slow-mode oscillations of hot
coronal loops. The events were observed with the SUMER spectrometer with a slit
located at the apex of a hot coronal loop which was visible in soft X-rays, but not seen
by EIT 195\AA. The spectra were recorded in a wide spectral window, including the lines
with formation temperatures from 10$^4$K to 10$^7$K. Two recurring brightening events 
associated with Doppler shift oscillations were observed only in the flare line 
Fe XIX (6.3 MK). The Doppler shift oscillations are strongly damped.
From the measured loop length and period, they estimated the phase speed for the
fundamental mode to be 240$-$380 \kms. The values are close to the sound speed at
6 MK, thus suggesting an interpretation in terms of global slow-mode oscillations.
Figure~\ref{sxrdp} demonstrates another example. Similarly, two recurring events of
Doppler shift oscillations were observed in the Fe XIX line. The initiation
of both events in association with a footpoint brightening suggests that
the loop oscillations could be triggered by small or micro-flares \citep{wan03b}. 

\begin{figure}
\centering
\includegraphics[width=0.9\linewidth]{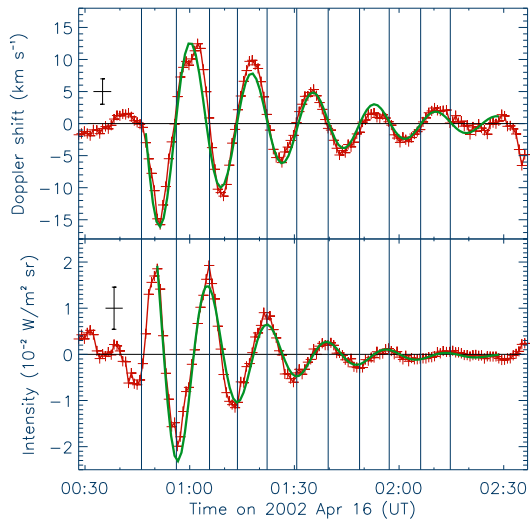}
 \caption{\label{waves}
{\em Top panel)} Evolution of Doppler shift and {\em bottom panel)} of line-integrated
intensity in the Fe XIX line at a distinct region of coherent
oscillations along the slit. The green curves are the best fits
with a damped sine function. The background shift and the
background intensity have been removed, respectively. \citep[from][]{wan03a} }
\end{figure}

Compressive slow-mode waves are expected to be associated with intensity
oscillations.  Figure~\ref{waves} shows one of the clearest examples.
The Doppler shift and line intensity oscillations are found not only with
the same period, but also having a phase shifted by exactly a 1/4-period, 
providing definitive evidence for a compressive standing wave in the loop
\citep{wan03a}. This is in contrast with a propagating compressive wave,
which shows velocity and intensity oscillations in phase.

Observations with a high cadence of 50 s made in April-May 2002 disclosed
phase propagation along the slit in some Doppler shift oscillations. The
measured propagating speeds cover a range of 8$-$102 \kms, with a mean
of $43\pm25$ \kms. \citet{wan03b} suggested that this feature may be due to
the oscillation triggered in a loop with fine structure or in a loop system. 

\begin{figure*}
\centering
\includegraphics[width=0.85\linewidth]{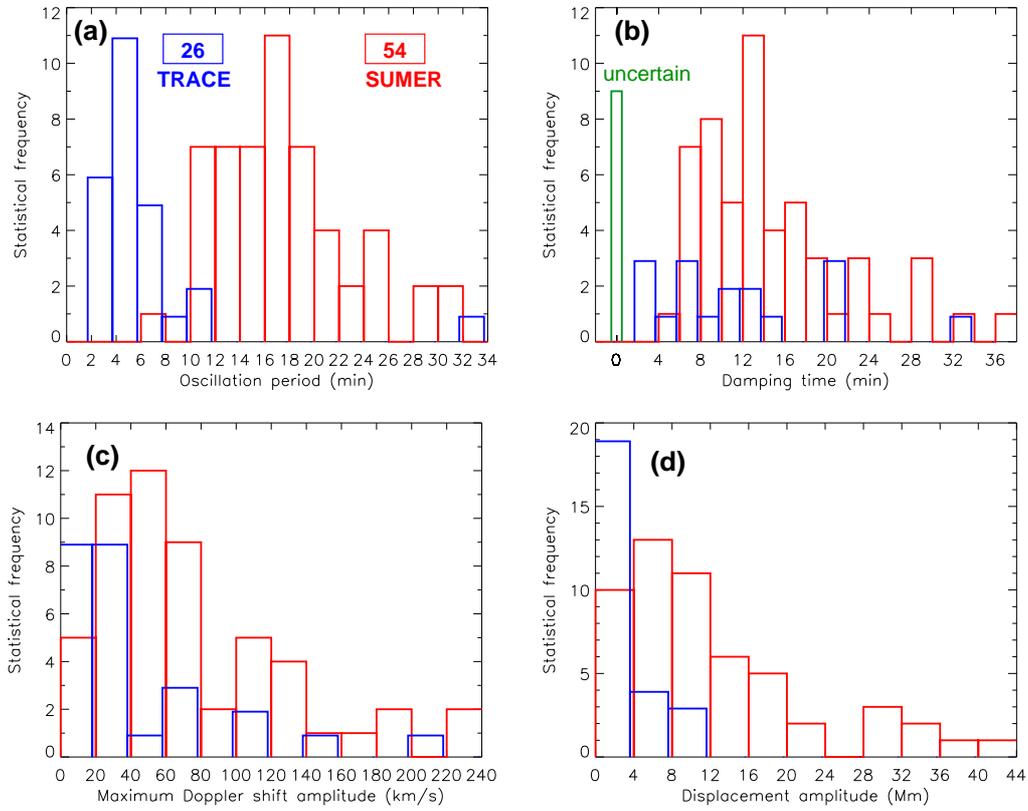}
 \caption{ \label{histo}
Comparison of the physical parameters for the 54 SUMER loop
oscillations and the 26 TRACE transverse loop oscillations.
(a)~Oscillation periods. (b)~Decay time. (c)~Velocity amplitude.
(d)~Displacement amplitude. \citep[from][]{wan03b} }
\end{figure*}

Figure~\ref{histo} shows that the histograms of physical parameters measured for 
the 54 SUMER oscillations \citep{wan03b} are compared with the result
for 26 TRACE transverse loop oscillations \citep{asc02}.
The SUMER loop oscillations have periods in the range 7$-$31 min
with a mean of 17.6$\pm$5.4 min, distinctly larger than those for the TRACE
case. The decay times in the range 6$-$37 min with a mean of
14.6$\pm$7.0 min, have a ratio to the period close to 1, which is about a factor 
of 2 shorter than the TRACE case. The velocity amplitude is comparable.
But the derived displacement amplitude is more than 5 times larger than
that measured from the TRACE transverse oscillations. These comparisons
also indicate that SUMER has detected a different oscillation mode from that 
found by TRACE.

Figure~\ref{twocas} shows a comparison of two brightening events. In one case, the slit
was located at the loop leg, and in the other case the slit at the loop apex.
The Doppler shift oscillations are clearly seen at the loop top, but
very weak in the loop legs. For almost all cases when the SXT observations are
available, it is found that the evident Doppler shift oscillations occurred at
the loop top \citep{wan03b, inn04}. This fact also supports that the hot 
oscillations observed by SUMER are the fundamental mode.

\subsection{Trigger and Excitation of Oscillations}
For all 54 cases, the initially rapid increase of the line intensity
and initial large Doppler broadenings indicate that the oscillations are 
excited impulsively. The initiation of some events was associated with the footpoint
brightening of the oscillatory loop (see Figs.~\ref{sxrdp}a-b), suggesting that 
the trigger could be small or micro-flares. \citet{wan05a} found that the spectral 
evolution often reveals the presence of two components in the initial phase of 
oscillations. This signature suggests that slow mode oscillations could be triggered
by hot plasma injection or energy release near a footpoint in the loop.
For 26 of 54 cases, they measured the highly shifted component of the 
Doppler shift on the order of 100$-$300 \kms. 

Figure~\ref{coolev}a shows an example for the hot plasma cooling in an oscillating 
loop from the temperature of more than 6 MK to 1 MK, as clearly evidenced from time 
delays of brightening subsequently seen in the lines Fe~XIX, Fe~XVII, Ca~XIII, and
Ca~X. The Ca~XIII emission shows a dimming when the Fe~XIX brightening starts,
indicative of an impulsive heating of the loop plasma. Figure~\ref{coolev}b shows 
the presence of Doppler shift oscillations simultaneously in the
Fe~XIX and Fe~XVII lines. Note that the strong initial redshifts are seen
in Fe~XIX, but invisible in Fe~XVII for both events, indicating that the initial
redshifts are not caused by the motion of local plasma in the loop, but by a pulse of
injected hot flow, which is possibly produced by the energy release at a footpoint.
The distinct difference of amplitudes of Doppler shift oscillations in these two
lines may indicate the oscillating plasmas initially at different temperatures.
Later on, their amplitudes gradually reach unanimity suggesting the arrival of
isothermal status.

In relation with simulations of the excitation of global slow mode standing waves in 
hot loops observed by SUMER, the following properties should be emphasized 
\citep{wan05a}:\\
1) Standing waves are set up quickly about a half period after the onset of the events.\\
2) These oscillations are the fundamental mode; no evidence for the second
harmonics are found yet.\\
3) Initial loop temperature is above 2$-$3 MK, then impulsively heated to a temperature
of 6$-$8 MK.\\
4) The duration of flarelike brightenings is several times the oscillation period.\\
5) Except for the strong initial injected hot flows lasting for about half a wave
period, no background flow is present during the event.

Based on 1D MHD simulations, some recent studies have explored the excitation
of slow-mode standing waves in coronal loops.  \citet{nak04b} and \citet{tsi04} 
show that the second harmonics can be excited in high temperature (30$-$40 MK) flare 
loops by impulsive energy deposit, independent of the
location. Whereas \citet{sel05b} show that the excitation of the fundamental mode
or the second harmonics depends on the location of the trigger. A hot pulse launched
at the loop apex excites the second harmonics, while a pulse at the loop leg or
footpoint excites the  fundamental modes. They also shows that the excitation time
of the standing wave at least 3 periods. \citet{tar05} show that the fundamental
mode can be set up immediately after the impulsive heat deposition at the
footpoint of a loop if the duration of the pulse matches the wave period.

\subsection{Damping Mechanism of Slow Mode Standing Waves}
Different from the damping mechanisms of fast kink mode oscillations such as resonant
absorption and shear viscosity, the damping of slow mode waves is mainly due to
thermal conduction and compressive viscosity. 
For typical parameters of the hot loops observed by SUMER, one-dimensional,
non-linear MHD simulations by \citet{ofm02c} show that the
large thermal conduction due to high temperature of the loop can lead to the
rapid damping of slow waves, and the expected scaling of the dissipation
time with period agrees well with the observations. Based on a 1D loop model with 
gravity stratification, \citet{men04} find that stratification 
results in further 10$-$20\% reduction of the wave-damping time.
This effect is not important because the pressure scale height at 6-8 MK is very 
large (about 300$-$400 Mm), compared to the height of typical coronal loops.

\begin{figure*}
\centering
\includegraphics[width=0.9\linewidth]{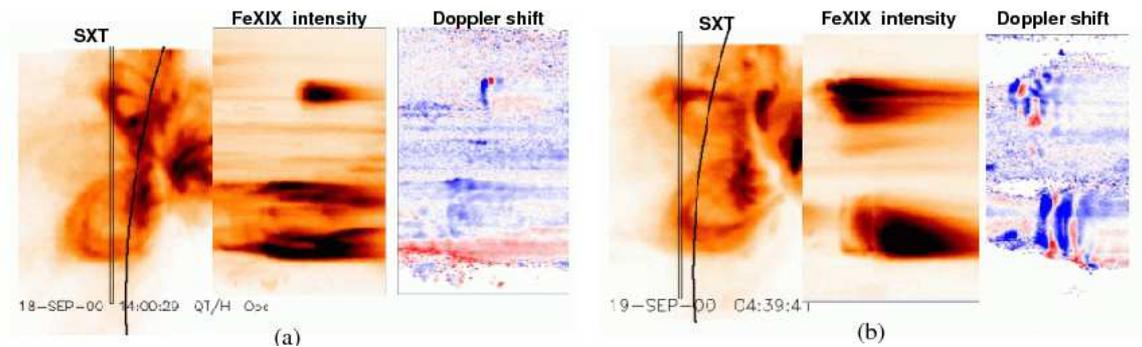}
 \caption{\label{twocas}
  Comparison of two brightening events observed with the SUMER slit
(a) located at the loop legs and (b) at the loop apex.  }
\end{figure*}

\begin{figure*}
\centering
\includegraphics[width=0.9\linewidth]{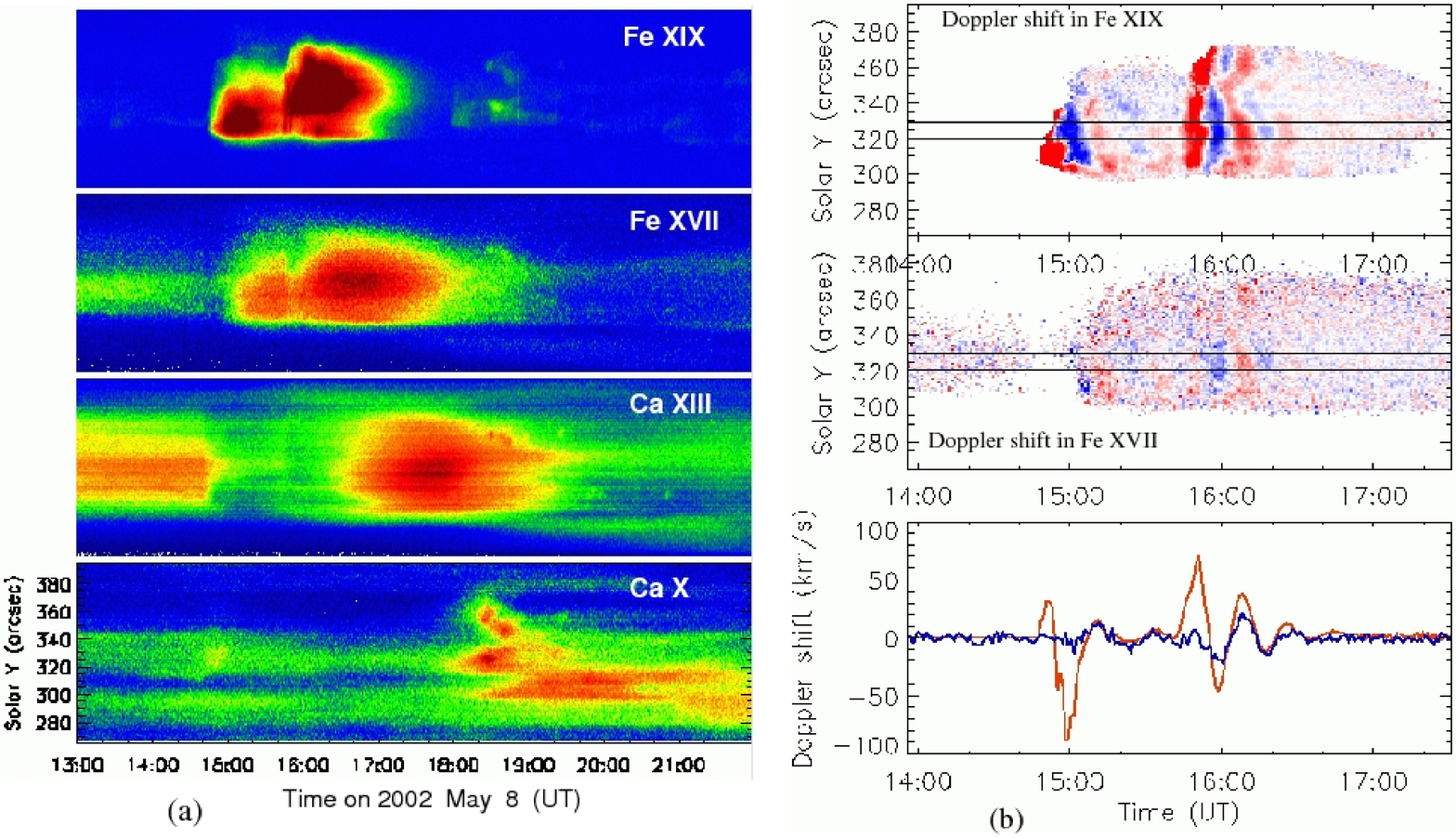}
 \caption{\label{coolev}
  (a) Time series of the line intensity along the slit for 4 spectral lines,
indicative of hot plasma cooling in the oscillation events observed with
SUMER. (b) Time series of Doppler shift along the slit seen in Fe XIX and
Fe XVII lines. The bottom panel shows the time profiles at a selected cut. 
The Doppler shifts in Fe XIX (Fe XVII) are represented by a red (blue) curve. }
\end{figure*}

\section{Propagating slow magnetoacoustic waves}

\subsection{Observations of Waves in Polar Plumes and Coronal Loops}

Periodic density fluctuations (period$\approx$ 9 min) in coronal plumes were 
first detected high above the limb by \citet{ofm97, ofm98,
ofm00b} using white light channel of the SOHO/UVCS. The ratio of the wave
amplitude in intensity to the background value grows with height \citep{ofm99}.
\citet{def98} detected quasi-periodic, propagating intensity disturbances
in EIT/SOHO observations of polar plumes. The periods are about $10-15$ min, 
and the propagating speeds are about $75-150$ \kms.  \citet{ofm99, ofm00a} 
found that these compressive disturbances could be modelled as slow 
magnetoacoustic waves.

\begin{figure}
\centering
\includegraphics[width=0.9\linewidth, height=1.2\linewidth]{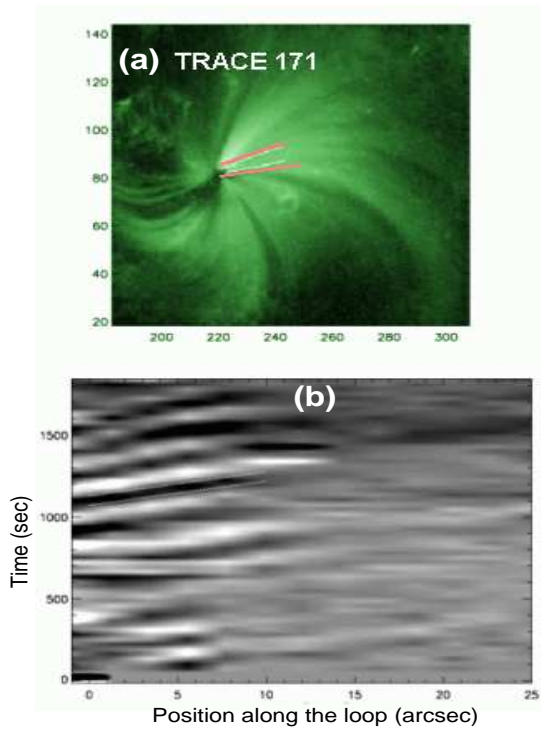}
 \caption{\label{prpwv}
  (a) Typical example (TRACE 171 \AA~$-$13 June 2001, 0646 UT) of a large
coronal loop footpoint supporting an oscillatory signal. (b) A plot of the
running difference between the average time series for each position along
the structure. \citep[from][]{dem02a} }
\end{figure}

\citet{ber99} first reported on the detection of similar propagating 
oscillations observed in coronal loops in SOHO/EIT 195 \AA. 
The propagating speed is of the order of 150 \kms.
Intriguingly, these propagating disturbances are only detected in fan of
straight loops, which are either widely open field lines or large scale loops 
connecting with the surrounding areas of the involved active region. 
\citet{dem00} confirmed this discovery with 
TRACE 171 \AA. Figure~\ref{prpwv} shows such a typical example. 
\citet{dem02a} found that loops that are situated above sunspot regions
display quasi-periodic intensity disturbances with a period of the order
of 3 min. whereas disturbances in ``non-sunspot'' loops with periods of
the order of 5 min.  These propagating disturbances in coronal loops
have been interpreted in terms of slow magnetoacoustic waves based on 
the MHD modelings \citep{nak00, tsi01}. 

Geometric and physical parameters of longitudinal oscillations were measured
by \citet{dem02b, dem02c} in 38 examples of large fan-like coronal loops
observed with TRACE 171 \AA~bandpass. They found that the disturbances travel
outward with a propagation speed of the order of 122$\pm$43~\kms and periods
of the order of 282$\pm$93~s. The amplitude
of the intensity variations are roughly 4.1$\pm$1.5\% of the background loop
brightness. The length of the examined loop footpoints is found to be 
26.4$\pm$9.7~Mm. The propagating disturbances are damped very quickly
and are typically only detected in the first 8.9$\pm$4.4 Mm.

\citet{rob01} found that the waves propagate in the same loop with different
phase speeds measured at 171\AA~and 195 \AA~bands, implying the temperature
structure in the loop. \citet{kin03} found that time series of propagating 
disturbances observed in the 171\AA~and 195 \AA~bands are highly
correlated, but with a tendency to decrease with distance along the structure.
\citet{mar03} analysed propagating oscillations with TRACE 171 \AA~images and
SOHO/CDS data in the He I, O V, and Mg IX lines, and found the presence of 
the waves in the different levels from the chromosphere to the corona.

\subsection{ Excitation of Longitudinal Oscillations}
\begin{figure}
\centering
\includegraphics[width=0.9\linewidth, height=1.1\linewidth]{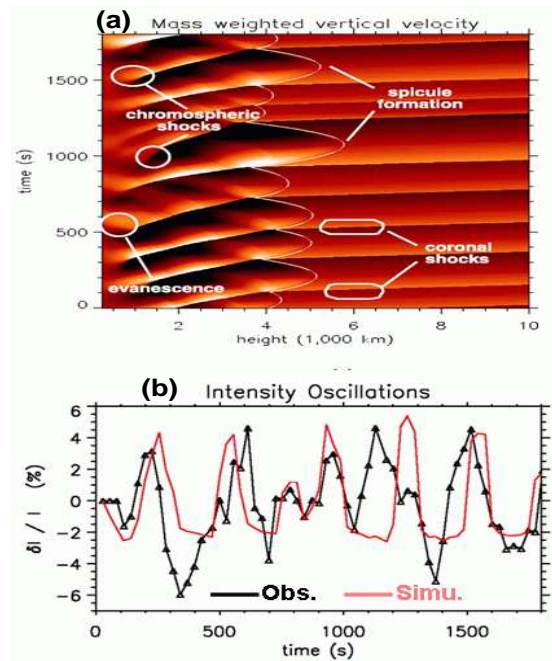}
 \caption{\label{simwav}
(a) Simulated plasma velocity along a rigid flux tube inclined by 40$^{\circ}$ 
from the vertical,
as a function of height and time (white is up, and black is down, weighted by density
for good visibility throughout the height range). (b) Running difference ($\delta{I}$) 
of loop intensity oscillations at one location (relative to total intensity I) as observed
in TRACE 171 Å (black) on 2001 June 12 from 7:25 UT onward (case 16a of De Moortel et al.
2002b). Shown for comparison are the results of a simulation (red) of a flux tube
(inclined by 35$^{\circ}$) driven by photospheric velocities observed at the footpoint of the loop.
\citep[from][]{dep05}}
\end{figure}

Since all these loops are quiescent and stable and the propagating disturbances
can remain for several consecutive hours, they are not flare-driven but are
most likely caused by an underlying driver exciting the loop footpoints
\citep{dem02a, dem02b, dem02c}. Many observations have shown 3-min period
oscillations in the chromosphere and transition regions above the sunspots, so in
those loops situated above the sunspots, the leakage of 3-min oscillations is an
explanation for the propagating slow waves in the coronal loops.  It is well known that
generally the 5-min p-mode oscillations cannot penetrate through the
chromosphere due to the acoustic cutoff. However, some recent work by \citet{dep04, dep05}
show that if the magnetic field near the loop footpoint has a large
inclination to the vertical, the increase of the acoustic cutoff period will allow the
leakage of 5 min p-mode oscillations into the corona. Figure~\ref{simwav}a shows 
the result from their 1D simulations.  It can be seen that the
photospheric oscillations develop into shocks which drive chromospheric spicules
and reach the corona, forming propagating slow magnetoacoustic waves observed by TRACE.
Figure~\ref{simwav}b shows a good agreement between the observed intensity oscillation 
and the simulated one which is driven by photospheric velocity oscillations observed by 
MDI. However, since their model does not include thermal conduction and radiative loss,
whether the damping scale of the coronal shocks in the loop consistent with the
observation or not needs further verification.

\subsection{Damping of Propagating Intensity Oscillations}

By modelling the propagation and dissipation of slow magnetoacoustic waves
in polar plumes, \citet{ofm00a} found that the nonlinear steepening of
the waves leads to enhanced dissipation owing to compressive viscosity
at the wave fronts and thus leads to damping of the waves within the first
solar radii.

\citet{nak00} modelled the propagation of slow waves in long coronal loops,
and found that the main factors influencing the wave evolution are dissipation
and stratification. Series of papers by \citet{dem03, dem04a, dem04b} investigated
various kind of effects which could contribute to the damping of slow waves, such as
gravitational stratification, field line divergence, coupling of slow and fast modes,
phase mixing of slow waves due to the horizontal density inhomogeneity, and showed that
the thermal conduction is the dominant damping mechanism.
Based on a more realistic model, Klimchuk et al. (2004) show that
thermal conduction, pressure and temperature stratifications are the most important
factors in the low corona to explain the observed damping of the intensity oscillations.
They conclude that the enhanced compressive viscosity is not necessary.

\section{Application of the MHD coronal seismology}
Coronal seismology has been used for determining physical parameters of coronal loops
in many studies. For example, based on rapid damping of the observed fast kink mode oscillations
by phase mixing, \citet{nak99, ofm02b} derived the value of the viscosity at least
5 orders of magnitude higher than the classical coronal value. While the 
interpretation of the damping by resonant absorption provided a new diagnostic 
of the density contrast of oscillating loops from the number of oscillation periods
\citep{asc03b}. Using the loop length and the period measured for the
observed global fast kink mode oscillations, \citet{nak01, asc02} obtained the mean value of
magnetic field strength in coronal loops. From a detection of multiple
harmonic standing transverse oscillations by \citet{ver04}, \citet{and05b} obtained
information about the density scale height in loops, based on the ratio of the period of
the fundamental mode to the period of the second harmonic differing from 2 in loops
with longitudinal density stratification.  Here, I present two examples to show
how the magnetic field strength can be determined from observations of the global
fast kink mode and global slow mode oscillations of coronal loops.

\subsection{Determination of the Magnetic Field Strength from Fast Kink Mode Oscillations}

Based on the theory of MHD modes of a straight magnetic cylinder \citep{edw83, rob84},
the period $P$ for the fundamental mode of fast kink oscillations is
\begin{equation}
 P=\frac{2L}{c_k},  \label{eqper}
\end{equation}
\noindent
where $L$ is the loop length and $c_k$ is the kink speed which is given by
\begin{equation}
c_k\approx{v_A}\left(\frac{2}{1+\rho_e/\rho_0}\right)^{1/2},   \label{eqck}
\end{equation}
\noindent
where $v_A=(B^2/4\pi\rho_0)^{1/2}$ is the Alfv\'{e}n speed in the loop, and
$\rho_{0}$ and $\rho_{e}$ are the density inside and outside the loop, respectively.
From Eqs.~(\ref{eqper}) and (\ref{eqck}), the value of the magnetic field in
the loop can be estimated by
\begin{equation}
B=(4\pi\rho_0)^{1/2}v_A=\frac{L}{P}\sqrt{8\pi\rho_0(1+\rho_e/\rho_0)}.
\end{equation}

Using the physical parameters measured from the 14 July 1998 event
\citep{nak99} and the 4 July 1999 event \citep{sch00}, \citet{nak01}
estimated the magnetic field in the range from 4~to~30 G.
Using the average values of $L$, $P$ and the lower limit to $n_{loop}$ (the loop
density) measured for 26 loop oscillations, \citet{asc02} obtained the
following scaling,
\begin{equation}
 B=18\left(\frac{L}{200~Mm}\right)\left(\frac{P}{300~s}\right)\sqrt{\frac{n_{loop}}{10^9~cm^
{-3}}}[G].
\end{equation}
\noindent
Considering each of the parameters has a variation by a factor of about 0.5,
they estimated the magnetic field values in a range of $B\approx3-30$ G,
which is consistent with the result obtained by \citet{nak01}.

\subsection{Determination of the Magnetic Field Strength from Slow Mode Oscillations}

From observations of the standing slow mode oscillations, the magnetic field strength 
of coronal loops  can also be estimated. In an example of the oscillating
loop (the lower one shown in Fig.~\ref{twocas}b), I measured the oscillation period
$P=17$ min and loop length $L=200$ Mm. The period for the fundamental mode of standing
slow mode oscillations is
\begin{equation}
   P=\frac{2L}{c_t},  \label{eqpslw}
\end{equation}
\noindent
where $c_t$ is the tube speed which is given by,
\begin{equation}
  c_t=\left(\frac{1}{c_s^2}+\frac{1}{v_A^2}\right)^{-1/2}, \label{eqct}
\end{equation}
\noindent
with $c_s$ the sound speed, given by
\begin{equation}
c_s=\left(\frac{\gamma{k_B}T}{\mu{m}_p}\right)^{1/2}=1.52\times10^4T^{1/2}, \label{eqcs}
\end{equation}
where $k_B$ is the Boltzman constant, $\mu$ is the mean molecular weight 
($\mu\approx{0.6}$ considering the He abundance in the corona) and $T$ is the temperature. 

With Eq.~(\ref{eqpslw}), the tube speed is obtained, $c_t=390$ \kms. In addition, the loop
temperature and electron density can be measured from the Yohkoh/SXT data using the 
filter ratio method \citep{tsu91}. In this case, $T\approx{8}$ MK and 
$n_0\approx{4}\times10^9$ cm$^{-3}$. With these parameters, I obtain the sound speed, 
$c_s=430$ \kms and Alfv\'{e}n speed $v_A=926$ \kms with Eq.~(\ref{eqct}). Finally I obtain
the magnetic field strength by $B=(4\pi\rho_0)^{1/2}v_A=4.6\times10^{-12}v_A n_0^{1/2}$, 
which is about 27 G, consistent with the results obtained from the fast kink mode 
oscillations. A statistical study with more than ten cases are in preparation by
\citet{wan05b}.

\section{Conclusions}
The main results are summarized in the following,\\
1) TRACE has discovered transverse (horizontal and vertical) loop oscillations, which
are interpreted as global fast kink mode oscillations. The exciter could be the blast wave
originating from the flare.\\
2) SUMER has discovered global slow-mode standing waves in hot ($>6$ MK) coronal loops.
The trigger could be small flare-like events at a single footpoint.\\
3) EIT and TRACE have found propagating intensity oscillations in fan-like coronal
loops, which are interpreted as the propagating slow magneto-acoustic waves. The
exciter may be the leakage of p-mode oscillations.\\
4) The observed kink and slow mode standing oscillations can be used as a new
diagnostic tool for determining the mean magnetic field strength in coronal loops. \\
5) These oscillations are all strongly damped. The fast kink modes are most likely
damped by resonant absorption and phase mixing, and the slow modes oscillations and
waves are mainly damped by thermal conduction.

With the future missions like Solar-B and Solar Orbit, detections of
high-frequency fast mode (kink and sausage) and torsional Alfv\'{e}n waves and
oscillations in coronal loops are expected, which are theoretically predicted to 
contribute to the heating of the corona.

\end{document}